\documentclass[prl,reprint,amsmath,amsfonts,amssymb,aps,superscriptaddress]{revtex4-1}
\usepackage{epsfig}
\usepackage{graphicx}  
\usepackage{verbatim} 
\usepackage{color}   
\usepackage{subfigure}
\usepackage{layouts} 
\usepackage{hyperref}  
\usepackage{bbm}
\usepackage{dsfont}

\newcommand{\Ham}{{\mathcal H}}

\newcommand{\btheta}{{\mbox{\boldmath$\theta$}}}

\newcommand{\bdelta}{{\mbox{\boldmath$\delta$}}}

\newcommand{\bS}{{\mathbf{S}}}

\newcommand{\bx}{{\mathbf{x}}}
\newcommand{\by}{{\mathbf{y}}}

\newcommand{\Like}{{\cal{L}}}
\newcommand{\LCDM}{{$\Lambda$CDM}}
\newcommand{\diff}{{\text{d}}}

\begin{document}

\title{Visualizing theory space: Isometric embedding of probabilistic
predictions, from the Ising model to the cosmic microwave background}
\author{Katherine N. Quinn, Francesco De Bernardis, Michael D. Niemack, James P. Sethna}
\affiliation{Physics Department, Cornell University, Ithaca, NY 14853-2501, United States}

\date{\today}

\begin{abstract}
We develop an {\em intensive embedding} for visualizing the space
of all predictions for probabalistic models, using replica theory.
Our embedding is isometric (preserves the distinguishability between models)
and faithful (yields low-dimensional visualizations of models with simple
emergent behavior). We apply our intensive embedding to the Ising 
model of statistical mechanics and the $\Lambda$CDM model applied to
cosmic microwave background radiation. It provides an intuitive, quantitative
visualization applicable to renormalization-group calculations and
optimal experimental design. 

\end{abstract}

\maketitle

What is the space of predictions available to a multiparameter model? How do we find the best parameters to fit experimental data? Which parameter combinations can be well estimated by the experimental data, and which change the predictions too subtly to be resolved?

For simple least-squares problems, where the model predictions are fit to data points $x_i$ with errors $\sigma_i$, these questions are naturally answered by the {\em model manifold} in data space, whose points are the model predictions $\by(\btheta)$ with coordinates given by the parameters $\btheta$. In this space parameter sets are considered close if their predictions are close, and so the best fit to data points $\bx$ is given by the closest point on the model manifold. The squared distance is given by the cost $\chi^2 = d^2(\by(\btheta),\bx)=\sum (y_i(\btheta)-x_i)^2/\sigma_i^2$. The rate at which the distance 
\begin{equation}
   d^2(\by(\btheta + \bdelta), \by(\btheta))
		\approx g_{\mu\nu} \bdelta^\mu \bdelta^\nu
\end{equation}
changes as parameters are shifted by a small distance $\bdelta$ distinguishes the stiff, easily measured combinations from the sloppy, ill-conditioned directions. Here the natural metric on the model manifold is induced from the embedding space, and is given by
\begin{equation}
\label{eq:DataSpaceEmbedding}
g_{\mu \nu} = \sum_i \frac{1}{\sigma_i^2}
		\frac{\partial y_i}{\partial \theta^\mu}
		\frac{\partial y_i}{\partial \theta^\nu}.
\end{equation}

\begin{figure}
	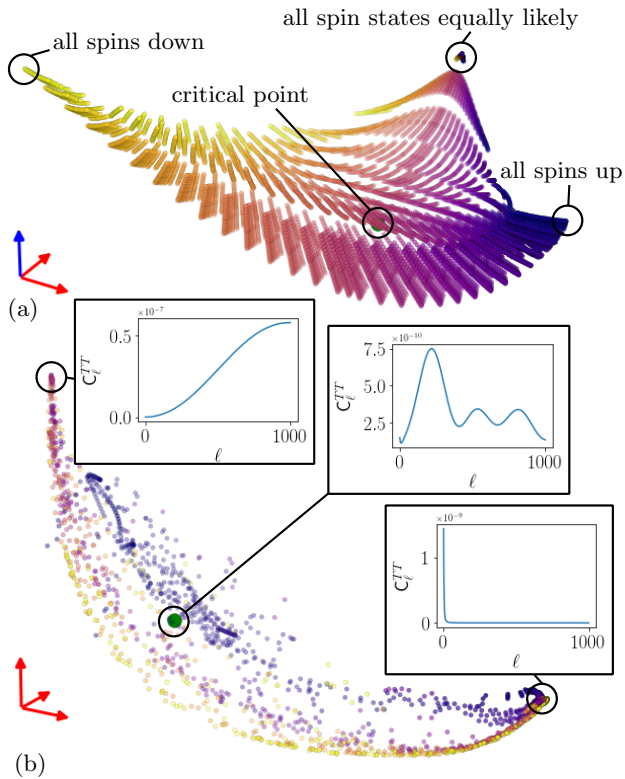
    \caption{{\bf Intensive model manifolds} for the Ising model of phase transitions in statistical mechanics and the \LCDM\ cosmological model predictions of the cosmic microwave background radiation. Each is a projection of the three largest principle components of the variations in model predictions. Unlike regular PCA decompositions, here certain directions correspond to positive (red axes) and negative (blue axes) square distances. (a)~The three-parameter $4\times 4$ Ising model, with external field $H$ (color), temperature (low to high pointing up toward the back peak), and a next-neighbor bond (varying `thickness'). The critical point for $h=0$ and $J=0$ is shown with a larger green dot, buried in the manifold along the central pink portion. The thickness involves the coordinate with negative squared distance, the other two projections are real. (b)~The six-parameter \LCDM\ model predictions, colored by $A_s$, the primordial fluctuation amplitude. Our universe is denoted by the large central point, and spectra are generated for temperature and $E$-polarization up to $\ell=1000$. The long direction mostly varies a combination of the $A_s$ and the optical depth at reionization (the scattering of light off of electrons during the epoch of reionization, when the first stars and galaxies ignited).}
    \label{fig:IntensiveManifolds}
\end{figure}

Although embedded in a large-dimensional space, the model manifold can be effectively projected for visualization using principle component analysis (e.g. Fig.~\ref{fig:IntensiveManifolds}). The fact that most physical systems show low-dimensional emergent behavior (a property which has been called `sloppiness'~\cite{SloppyReview})
make these projections useful.

In this manuscript, we provide a generalization of this embedding to models whose predictions are probability distributions. Models can be interpreted generally as providing a prediction for the likelihood $\Like(\bx|\btheta)$ of experimental data $\bx$ given $\btheta$. In the case of least-squares fitting, this is proportional to the exponential of minus half of the cost, $\Like(\bx|\btheta) \propto \exp(-\chi^2/2)$ .

We shall study two disparate models, the Ising model for phase transitions (Fig.~\ref{fig:IntensiveManifolds}~a) and the dark energy and cold dark matter (\LCDM) cosmological model (Fig.~\ref{fig:IntensiveManifolds}~b) predictions of the microwave background radiation. The generalized, long-bond Ising model
we study is discussed in~\cite{MachtaCTS13} and described by the Hamiltonian
\begin{align}
\label{eq:Ham}
\Ham_\btheta(\bS) =& h \sum_{i,j}s_{i,j} - J \sum_{i,j}\left(s_{i,j}s_{i+1,j} + s_{i,j}s_{i,j+1}\right)\nonumber \\
& - J'\sum_{i,j}\left( s_{i,j}  s_{i+1,j+1} +s_{i,j}s_{i+1,j-1} \right).
\end{align}

Here $\bS = \{s_{i,j}\}$ represents the spin configuration with $s_{i,j} = \pm 1$, and $\btheta = \{h,J,J'\}$ are the parameters representing the field, nearest and next-nearest neighbor coupling with the temperature set to one. The predictions of this Ising model are the Boltzmann probabilities of observing a particular spin configuration $\bS$, given by $\Like(\bS|\btheta)=\exp(-\Ham_\btheta(\bS))/Z$ where the partition function $Z$ serves to normalize the distribution to one.

The \LCDM\ model we study has six parameters, which can be described as the Hubble constant ($h_0$), the physical baryon density ($\Omega_bh^2$), the physical cold dark matter density ($\Omega_ch^2$), the primordial fluctuation amplitude ($A_s$), the scalar spectral index ($\eta$), and the optical depth at reionization ($\tau$). The \LCDM\ model predicts the angular power spectrum of temperature and polarization anisotropies in maps on the CMB, in addition to other cosmological observables. We use the CAMB software package~\cite{Lewis:1999bs} to generate power spectra for a range of \LCDM\ parameters in this analysis. Parameter ranges are detailed in~\cite[Table~SI]{SI}. To compare to the model, CMB maps are decomposed into spherical harmonics with coefficients $a_{\ell m}$. The angular power spectral of temperature and polarization anisotropies are computed as $C_\ell = \frac{1}{2 \ell +1}\sum_m \left|a_{\ell m}\right|^2$. Measurements of the $C_\ell$ amplitudes from telescopes on satellites, balloons, and the ground (e.g.~\cite{2013pss1.book..431H}) provide thousands of independent $C_\ell$ measurements from large angular scales (low $\ell$) to few arcminute angular scales ($\ell \approx 3000$) that are fit versus $\ell$ to the \LCDM\ model. The predictions of the \LCDM\ model can more generally be viewed as providing the probability of observing a sky map with a particular power spectrum, since the $C_\ell$ predicted by the theory represent the correlations and cross-correlations. The likelihood of a particular map given a set of cosmological parameters $\btheta$ is represented as:
\begin{align}
	\Like(\{a_{\ell m}\} | \btheta) = \prod_{\ell m}\frac{1}{\sqrt{(2\pi )^3 |C_l|}}\exp\left(-\frac{1}{2}a_{\ell m}^\dagger C_\ell^{-1}a_{\ell m}\right)
\end{align}

In the least-squares problem, the natural metric between two nearby predictions is given by the Euclidean distance between points in the data space. For general probability distributions, the natural metric is the Fisher information
\begin{equation}
\label{eq:FisherInformation}
g_{\mu\nu} = \int \diff \bx \frac{\partial \log \Like (\bx| \theta )}{\partial \theta^\mu} \frac{\partial \log \Like (\bx| \theta )}{\partial \theta^\nu} \Like(\bx|\theta).
\end{equation}
As the parameters $\btheta$ are varied, neither the embedding $\Like(\bS|\btheta)$ into the $2^N$ dimensional spin probability simplex for the Ising model nor the predictions $C^{XY}_{\ell}$ in the space of variances~\cite[Sec.~I,~II]{SI} forms an isometric embedding; both distort the natural local distances between model predictions.

\begin{figure}
        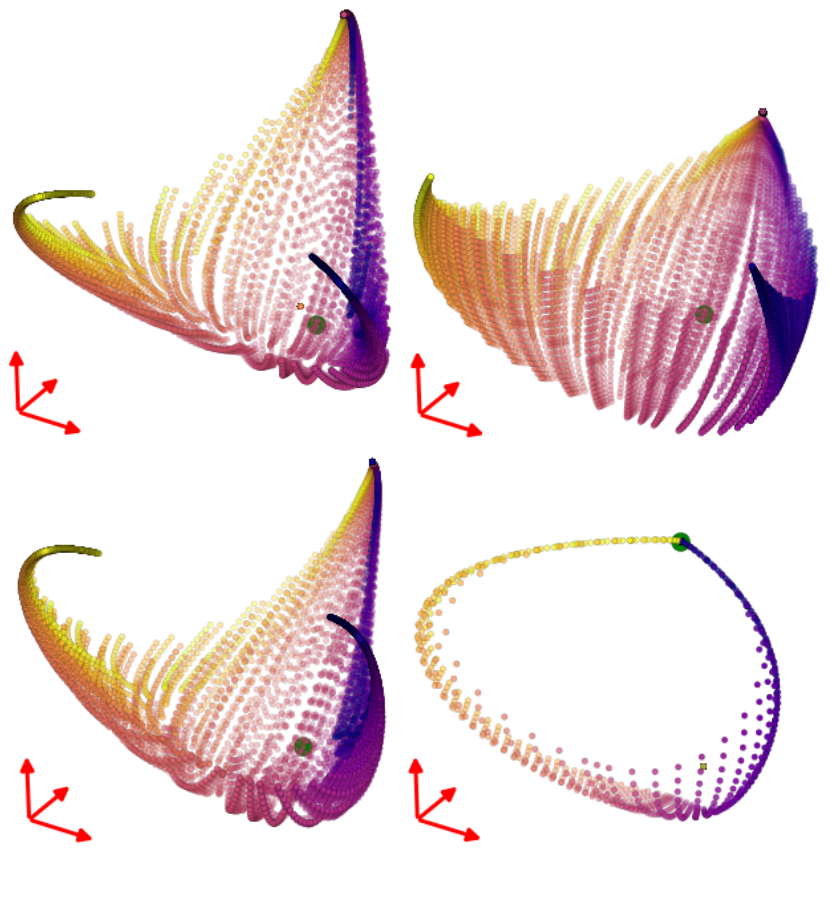
    \caption{{\bf Extensive model manifolds} for the Ising model and replicas thereof. In all images, the critical point is indicated by a large point. (a)~The three-parameter $4\times 4$ Ising model, embedded on the $2^{16}-1$ dimensional sphere using the square root of the Boltzmann probability for each spin configuration. (b)~The $2\times 2$ Ising model. Note that the predictions are less `curled up' at the edges. (c)~Four replicas of a $2\times 2$ Ising model, providing the same amount of data as the $4\times 4$ Ising model, and showing symptoms of curling around the sphere. (d)~The $2\times 2$ Ising model, replicated 512$^2$ times to mimic a traditional $1024\times1024$ Ising model. Small changes in parameters lead to orthogonal probability distributions, located ninety degrees apart on the $n$-sphere.}
    \label{fig:ExtensiveManifolds}
\end{figure}

There is a straightforward isometric embedding~\cite{hellinger,wiki:FIM,Gromov13}, which is unfortunately unusable in practice for our purposes. One may check~\cite[Section S-II.A]{SI} that $2\sqrt{\Like(\bx | \btheta)}$ embeds the predictions of a general probabilistic model onto the positive octant of an $n$-sphere of radius 2 in a way that preserves the metric tensor given by the Fisher information, Eq.~\ref{eq:FisherInformation}. The dimension of the $n$-sphere is given by the dimension of the data, and may be finite or infinite. We denote the cosine of the angle between the predictions of two parameter sets $\btheta_1$ and $\btheta_2$ on the model manifold as
\begin{align}
\label{eq:DotUnreplicated}
\left<\btheta_1 ; \btheta_2\right> &= \left\langle \sqrt{\Like(\bS|\btheta_1)},\sqrt{\Like(\bS|\btheta_2)}\right\rangle\nonumber\\
&= \sum_\bx \sqrt{\Like(\bx|\btheta_1)} \sqrt{\Like(\bx|\btheta_2)}
\end{align}
(just 1/4 the dot product of the two points on the $n$-sphere of radius two,
hence the notation).
The squared distance in the embedding space is thus
\begin{align}
\label{eq:dUnreplicated}
&d^2(\btheta_1,\btheta_2)  \nonumber\\
&=\left\langle 2\sqrt{\Like(\bx|\btheta_1)} - 2\sqrt{\Like(\bx|\btheta_2)} ,
   2\sqrt{\Like(\bx|\btheta_1)} - 2\sqrt{\Like(\bx|\btheta_2)} \right\rangle \nonumber\\
&= 8 \left(1- \left\langle\sqrt{\Like(\bx|\btheta_1)},\sqrt{\Like(\bx|\btheta_2)}\right\rangle\right) \nonumber \\
&= 8 \left(1-\left<\btheta_1;\btheta_2\right>\right)
\end{align}

For small system sizes, such as the $4\times4$ Ising model, this is a useful isometric embedding and shown in Fig.~\ref{fig:ExtensiveManifolds}~a. When increasing system sizes, such as considering a larger Ising model or expanding CMB spectra to a high order in spherical harmonics, even slightly different sets of parameters will lead to easily distinguishable predictions in the sense that any prediction whose probability is large for one set of parameters will be very unlikely for the other. Hence the dot products become approximately zero, leading to the the models predictions becoming 90$^\circ$ apart on the $n$-sphere of radius two, as shown in Fig.~\ref{fig:ExtensiveManifolds}. For our exploration of the possible CMB spectra, most pairs of parameters very neatly the maximum possible distance of $\sqrt{8}$.~\cite[Fig.~SI]{SI}.

What we want is an embedding which unwinds the manifold, revealing its $intensive$ properties (information per data point, rather than total information). As a motivation for this approach, consider how four spin configurations drawn from an $L\times L$ Ising model roughly gives us the same information \footnote{Unless the system is near a critical point, where the correlations between spins extend beyond the size of the system.} about the parameters as would one configuration drawn from a $2L\times2L$ Ising model since each of these systems has the same number of neighboring pairs, the same number of next-neighbor pairs, {\em etc.} We can therefore reproduce the orthogonality catastrophe of embedding large Ising models by considering multiple replicas of small Ising models, see Fig.~\ref{fig:ExtensiveManifolds}~c and Fig.~\ref{fig:ExtensiveManifolds}~d; the $N$-times replicated $2\times2$ Ising model stretches and wraps around the $n$-sphere, soon hiding the useful geometrical information about model predictions as the number of replicas goes to infinity as show in Fig.~\ref{fig:ExtensiveManifolds}~d.

What we want instead is the limit as the number of replicas goes to \textit{zero}. Just as the energy of a large system is extensive (growing with system size) and the temperature is intensive (independent of size), we want an intensive manifold embedding that reflects the information density about the parameters embodied in the likelihood $\Like(\bx|\btheta)$, not the total information. If the cosine angle for the unreplicated model is 
$\left\langle\sqrt{\Like(\bx|\btheta_1)}, \sqrt{\Like(\bx|\btheta_2)}\right\rangle$, what is it for the replicated model? The likelihood of finding data $\bx_1,\dots,\bx_N$ for a given parameter set $\btheta$ is 
\begin{equation}
\label{eq:ReplicatedLike}
\Like^{(N)}(\{\bx_1,\dots,\bx_N\}|\btheta) = 
  \Like(\bx_1|\btheta)\cdots \Like(\bx_N|\btheta).
\end{equation}

Similarly to the non-replicated cosine between points from Eq.~\ref{eq:DotUnreplicated}, we can define the cosine of the angle for this replicated system;
\begin{align}
\label{eq:DotReplicated}
&\left<\btheta_1^{(N)};\btheta_2^{(N)}\right> \nonumber \\
&= \left< \sqrt{\Like^{(N)}(\{\bx_1,\dots,\bx_N\}|\btheta_1)}, \sqrt{\Like^{(N)}(\{\bx_1,\dots,\bx_N\}|\btheta_2)}\right> \nonumber \\
&= \sum_{\bx_1} \dots \sum_{\bx_n}
\sqrt{\Like^{(N)}(\bx_1,\dots,\bx_N|\btheta_1)}
\sqrt{\Like^{(N)}(\bx_1,\dots,\bx_N|\btheta_2)} \nonumber\\
&= \left(\sum_\bx \sqrt{\Like(\bx|\btheta_1)} \sqrt{\Like(\bx|\btheta_2)}\right)^N \nonumber\\
&= \left<\btheta_1;\btheta_2\right>^N.
\end{align}
Hence the distance in the replicated model is
\begin{align}
\label{eq:dReplicated}
d^2_{(N)}(\btheta_1,\btheta_2)  
&= 8\left(1-\left<\btheta_1^{(N)};\btheta_2^{(N)}\right>\right) \nonumber \\
&= 8\left(1-\left<\btheta_1;\btheta_2\right>^N\right).
\end{align}

Now we are poised to take the limit as the number of replicas goes to zero, giving us an intensive model manifold embedding, and note that $\lim_{N\to 0} (Z^N - 1)/N = \log(Z)$.~\footnote{This is the `replica trick' used in the calculation of partition functions for disordered systems~\cite{MezardM09}} We define the intensive squared distance between two parameter predictions to be the squared distance per replica in the limit $N\to 0$, or
\begin{equation}
\label{eq:dIntensive}
d^2_{I}(\btheta_1,\btheta_2)  
= \lim_{N\to 0} -8\frac{\left(\left<\btheta_1;\btheta_2\right>^N-1\right)}{N}
= -8 \log \left<\btheta_1;\btheta_2\right>.
\end{equation}

The intensive metric distance between two predictions in Eq.~\ref{eq:dIntensive} is precisely the $\chi^2$ distance for the simple least-squares problem described early on. For example, a least-squares model $y(\theta) = \theta$ with error bar $\sigma$ has a natural Euclidean metric
\begin{equation}
\label{eq:d2residual}
d^2_\mathbf{E}(\theta_1,\theta_2) = 
	\sum_i (y_i(\theta_1)-y_i(\theta_2))^2/\sigma_i^2 
	= (\theta_1-\theta_2)^2/\sigma^2.
\end{equation}
Here, the likelihood function is simply a Normal distribution with mean $\theta$ and standard deviation $\sigma$, which we denote $\mathcal{N}(\theta,\sigma)$. This Euclidean metric is identical to the intensive embedding of the same model:
\begin{align}
\label{eq:d2I}
d^2_I(\theta_1,\theta_2)
    =& -8 \log \left<\btheta_1;\btheta_2\right> \nonumber \\
    =& -8 \log \left(\int \diff x \sqrt{\Like(\bx|\btheta_1)}\sqrt{\Like(\bx|\btheta_2)}\right) \nonumber\\
    =& -8 \log \left(\int \diff x \sqrt{\mathcal{N}(\theta_1,\sigma)} \sqrt{\mathcal{N}(\theta_2,\sigma)}\right) \nonumber \\
    =& (\theta_1-\theta_2)^2/\sigma^2.
\end{align}

The replica trick we use in Eq.~\ref{eq:dIntensive} has yielded fruitful insights and correct solutions in challenging problems in statistical physics~\cite{MezardM09,Parisi79,MezardPV87}; however, its applicability and interpretation is not mathematically well understood. Indeed, we find that the limit $N\to 0$ usefully avoids constraints like the triangle inequality by yielding an intensive manifold embedded in a zero-dimensional data space yet can be projected along multiple orthogonal directions, some of which have imaginary distances. Consider the $4\times4$ Ising model with $16$ spins that predicts a probability distribution for each of $2^{16}$ spin configurations, so the $n$-sphere embedding is a manifold in $\mathbb{S}^{2^{16}}$ shown in Fig.~\ref{fig:ExtensiveManifolds}~a. The $N$-replicated embedding is thus a manifold in $\mathbb{S}^{N 2^{16}}$, with dimension $N 2^{16}$ that goes to 0 as $N\rightarrow 0$.

Despite being zero-dimensional, our intensive manifold can nonetheless can be projected along any of several orthogonal directions (as shown in Fig.~\ref{fig:IntensiveManifolds} which depicts three-dimensional projection of the Ising and CMB manifolds). The standard method for extracting the `long' directions of a multidimensional data set is principal component analysis, which centers the data into columns $\by^{(J)}$ of a non-square matrix $M_{iJ} = y_i^{(J)}$, and then uses a singular value decomposition $M = U \Sigma V^T$. The coordinates of a data point $J$ in the orthonormal basis of columns of $V$ are given by $\Sigma_{\alpha \alpha} U_{J\alpha}$. But the squared singular values and the columns of $V$ are the eigenvalues and eigenvectors of the square matrix of dot products $(MM^T)_{JK} = \sum_i M_{iJ} M_{iK} = \by^{(J)}\cdot\by^{(K)}$. In our replica theory limit, it is precisely these dot products that we can extrapolate as $N\to 0$ (Eq.~\ref{eq:dReplicated}, see SI~\cite[S-IV]{SI}). However, unlike the positive definite eigenvalues of Wishart matrices (of the form $MM^T$), our intensive principal component limit can and often does have negative eigenvalues -- leading to imaginary components for the coordinates of data points in the principal component basis. Just as in Minkowski space for special relativity, the squared straight-line distance between the predictions of two parameter combinations for the intensive manifold embedding is given by the sum of the squares of the differences between these coordinates, and hence imaginary displacements correspond to negative contributions to this squared distance. In particular, in Fig.~\ref{fig:IntensiveManifolds}, the next-nearest neighbor direction for the Ising model contributes negative values to the squared distance.

The metric tensor $g_{\mu\nu}$ of Eq.~\ref{eq:FisherInformation} is preserved under any isometric embedding, and for every integer $N$ we have a genuine Euclidean isometric embedding, so we expect that our intensive manifold will preserve distances between nearby points. One can check in our mathematically problematical extrapolation $N\to 0$ that our intensive manifold does indeed preserve this metric~\cite[S-III]{SI}. In particular, nearby points have `space-like' positive intensive distances. However, as two probability distributions become orthogonal, their intensive squared distance $d_I^2(\btheta_1,\btheta_2) = -8\log\left(\sum_x \sqrt{\Like(x|\btheta_1)} \sqrt{\Like(x|\btheta_1)}\right)$ goes to infinity. For example, the intensive distance between two Ising models with external fields $h=\pm \infty$ will be infinite, since each has zero probability except in different fully magnetized states $s_\uparrow$ and $s_\downarrow$. (This is in contrast to the limit of $\sqrt{8}$ for the n-sphere embedding, but in agreement with the least-squares residual distance.) However, for finite fields (with non-zero temperature) there is a non-zero probability of being in a state with all spins up or down. Therefore, our intensive distance between these two extreme points and any other is finite. The triple ($h = \infty$,~$h=-\infty$,~$\btheta$) for any parameter set $\btheta$ therefore violates the triangle inequality (the sum of the lengths of two sides is smaller than the third). The Minkowski metric may be seen as the mathematical mechanism allowing the triangle inequality to be violated.

How do we envision using our intensive model embedding? Fig.~\ref{fig:IntensiveManifolds}~b illustrates the distinguishability within the six-parameter family of \LCDM\ models given recent CMB measurements. This approach could be used to guide the design of future instruments to measure new cosmological parameters, such as the neutrino mass sum, the number of relativistic species, and the tensor-to-scalar ratio. One could map out the model manifold, incorporating these new parameters into the model and the new measurement probes into the data space, and then take a cross section  to be consistent with current CMB data about our universe, allowing a nonperturbative characterization of the limits of parameter uncertainties. Fig.~\ref{fig:ExtensiveManifolds}, illustrating the family of behaviors exhibited by Ising models, could easily be coarse-grained by sampling a subgrid of spins in a large Ising model. The renormalization group tells us that this coarse-grained model can be rescaled to match the original model at renormalized parameters; distance in the intensive metric embedding could be a systematic, principled way of matching these parameters.

Understanding how to manipulate, distill, and visualize large data sets in high dimensions is a vast and sophisticated field, spanning mathematics, statistics, and machine learning. Our intensive manifold embedding promises to have wide applicability in interpreting Bayesian models of complicated systems, visualizing and distilling the statistical mechanics of phases and phase transitions, and guiding optimal experimental design in poorly constrained systems, despite the Minkowskian weirdness of the $N\to 0$ replica limit. Perhaps this weirdness is the reason that this manifestly useful geometrical construction was overlooked.

We thank Mark Transtrum for guidance on algorithms, Colin Clement for ideas on isometric embeddings, and both for useful conversations. KNQ was supported by a fellowship from the Natural Sciences and Engineering Research Council of Canada (NSERC), and JPS and KNQ were supported by the National Science Foundation through grant NSF DMR-1312160 and DMR-1719490. MDN was supported by NSF grant AST-1454881.

\bibliography{IntensiveManifold}

\end{document}


\beginsupplement

\title{Supplemental material for \break
Visualizing theory space: Isometric embedding of probabilistic
predictions, from the Ising model to the cosmic microwave background}
\author{Katherine N. Quinn, Francesco De Bernardis, Michael D. Niemack, James P. Sethna}
\affiliation{Physics Department, Cornell University, Ithaca, NY 14853-2501, United States}

\date{\today}

\begin{abstract}
\end{abstract}

\maketitle

This is the supplemental material accompanying {\em Visualizing theory space: Isometric embedding of probabilistic predictions, from the Ising model to the cosmic microwave background}.

Section~\ref{sec:OmegaMatrix} discusses the general non-Euclidean embedding provided by models whose errors are dependent upon parameter values, with particular reference to the cosmic microwave background correlation function where the fluctuations are the predictions of the model. Section~\ref{sec:HypersphereEmbeddingFailures} shows that $2\sqrt{\Like (\bx | \btheta)}$, two times the square root of the likelihood of a fit, acts as an isometric embedding for a general probabilistic model onto an $n$-sphere of radius two. Section~\ref{subsec:CMBHypersphere} illustrates the failures of this $n$-sphere embedding as a practical tool for models whenever the data provides good discrimination between different model predictions. (The proposed intensive embedding in the main text takes a formal limit of the embedding as the amount of data goes to zero to bypass this challenge.) Here we focus on the \LCDM\ model and cosmic microwave background (CMB) anisotropy predictions, complementing the discussion in the main text of the Ising model. Section~\ref{sec:IntensiveIsometricity} shows that our intensive manifold embedding is isometric -- faithfully representing the distances between predictions of nearby models, as characterized by the Fisher information. Finally, section~\ref{sec:IntensivePCA} describes in detail how one generalizes principal component analysis to the intensive manifold embedding, fleshing out the discussion in the main text by explicitly centering the model predictions before taking the limit of zero replicas and implementing the singular value decomposition.

\section{Cosmic microwave correlations as a non-Euclidean embedding}
\label{sec:OmegaMatrix}

The anisotropy in CMB radiation can be characterized by a $2\times 2$ direction dependent intensity matrix $I_{ij}(\hat{n})$ whose components can be recognized as 3 of the 4 Stokes parameters, \textit{I}, \textit{Q}, and \textit{U}. The \textit{Q} and \textit{U} polarization maps can be made independent of the Stokes parameter measurement basis by separating them into divergence (\textit{E}) and curl (\textit{B}) components, to generate three maps of cosmological interest; the temperature fluctuation map \textit{T} and two polarization maps, \textit{E} and \textit{B}. These can be expanded into spherical harmonics,
\begin{equation}
X(\hat{n}) = \sum_{\ell m}a_{\ell m}^X Y_{lm}(\hat{n})\quad\text{where}~X={T,E,B}.
\end{equation}
The anisotropies are expected to be (approximately) Gaussian. All of the Gaussian information can be extracted form the angular power spectra, which are defined defined as the cross correlation of the coefficients in the expansion and written as
\begin{equation}
C_\ell^{XY}\equiv\frac{1}{2\ell+1}\sum_{m}\left<a_{\ell m}^X a_{\ell m}^Y\right>\quad\text{where}~X,Y=T,E,B.
\end{equation}
Using this, we can construct a correlation matrix for the fluctuations,
\begin{equation}
C_\ell = \begin{pmatrix} C_\ell^{TT} & C_\ell^{TE} & C\ell^{TB} \\ C_\ell^{TE} & C_\ell^{EE} & C_\ell^{EB} \\ C_\ell^{TB} & C_\ell^{EB} & C_\ell^{BB}  \end{pmatrix}.
\end{equation}

The values of $C_\ell$ depend on the $\Lambda$CDM parameters, and likelihood analyses of CMB data fit with such a correlation function have  been extensively studied, as they are invaluable for fitting and forecasting CMB measurements~(e.g.~\cite{PlanckLikelihood,CMBHamimeche,CMBPolLikelihood,2016arXiv161002743A}). The probability of a fit for this data can be expressed as
\begin{equation}
p(\{\hat{a}_{\ell m}\}|\btheta) = \prod_{\ell m}\frac{1}{\sqrt{(2\pi)^3|C_\ell|}}\exp\left(-\frac{1}{2}\hat{a}_{\ell m}^\dagger C_\ell^{-1}\hat{a}_{\ell m}\right).\label{eq:probCMBDef}
\end{equation}
This conditional probability defines the likelihood~\citep{PlanckLikelihood,CMBPolLikelihood}, $\mathcal{L}(\{\hat{a}_{lm}\}|\btheta) = p(\{\hat{a}_{lm}\}|\btheta)$. The metric is given by the Fisher Information Matrix (FIM),
\begin{equation}
g_{\alpha\beta}(\btheta) = -\int\left(\partial_\alpha\partial_\beta\log \Like(\bx|\btheta)\right) \Like(\bx|\btheta)\diff\bx.\label{eq:secondDerFIM}
\end{equation}
We can evaluate this integral by looking at the second derivatives of $\mathcal{L}$;
\begin{eqnarray}
-\partial_\alpha\partial_\beta\log\Like(\{\hat{a}_{\ell m}\}|\btheta)&=&\frac{1}{2}\sum_{\ell m}\partial_\alpha\partial_\beta\left(\log|C_\ell|+\hat{a}_{\ell m}C_\ell^{-1}\hat{a}_{\ell m}\right)\nonumber\nonumber\\
&=&\frac{1}{2}\sum_{\ell m}\left(\frac{\partial_\alpha\partial_\beta|C_\ell|}{|C_\ell|}-\frac{\partial_\alpha|C_\ell|\partial_\beta|C_\ell|}{|C_\ell|^2}\right)\nonumber\\
&~&+\frac{1}{2}\sum_{\ell m}\hat{a}_{\ell m}\partial_\alpha\partial_\beta C_\ell^{-1}\hat{a}_{\ell m}.
\end{eqnarray}

This expansion can be combined with Eq.~\ref{eq:secondDerFIM} to extract all terms independent of the data. Thus, the first two terms in the sum can be completely pulled out of the integral. The remaining term is harder, and to evaluate it we make use of the following integral  for symmetric, positive definite $M\times M$ matrix $A$ and symmetric $M\times M$ matrix $B$
\begin{eqnarray}
\sqrt{\frac{|A|}{(2\pi)^M}}\int \textbf{x}^T B \textbf{x} \exp\left(-\frac{1}{2}\textbf{x}^T A \textbf{x}\right)d\bx = \text{Tr}(A^{-1} B).\nonumber\\
\end{eqnarray}

This allows us to solve Eq.~\ref{eq:secondDerFIM}, setting $A=C_\ell^{-1}$ and $B=\partial_\alpha\partial_\beta C_\ell^{-1}$. We can now combine all the pieces together, and obtain a formula for the FIM
\begin{eqnarray}
g_{\alpha\beta}(\theta) =&~&\sum_{\ell}\frac{2\ell+1}{2}\left(\frac{\partial_\alpha\partial_\beta|C_\ell|}{|C_\ell|} -\frac{\partial_\alpha|C_\ell|\partial_\beta|C_\ell|}{|C_\ell|^2} \right)\nonumber\\
&+&\sum_{\ell}\frac{2\ell+1}{2}\text{Tr}\left(C_\ell\partial_\alpha\partial_\beta C_\ell^{-1}\right).\label{eq:FinalFIM}
\end{eqnarray}

We can compare this to to previous results for FIM derivations,~\cite{CMBFisher,CMBPolLikelihood} and confirm that we obtain the same result. We can decompose this as a sum over $\ell$ and the different spectra to obtain:
\begin{equation}
g_{\mu\nu} = \sum_{\ell,XY,X'Y'}J^l_{XY,\mu}\Omega^l_{XY,X'Y'} J^\ell_{X'Y',\nu}= (J^T\Omega J)_{\mu\nu}\label{eq:FIMCMB}
\end{equation}
where $J^\ell_{XY,\mu}=\partial C^{XY}_\ell/\partial \theta_\mu$ is a tensor of partial derivatives, and $\Omega$ is given by Eq.~\ref{eq:FinalFIM} where derivatives are taken with respect to the $C_\ell^{XY}$. We can express $\Omega$ as a block diagonal matrix because the $C_\ell^{XY}$ are uncorrelated for different values of $\ell$, and in such a form represents the FIM where the parameters of interest are the $C_\ell^{XY}$. Note that $g_{\mu\nu}$ is not constant in the coordinates given by $C_\ell$.

\begin{figure*}
        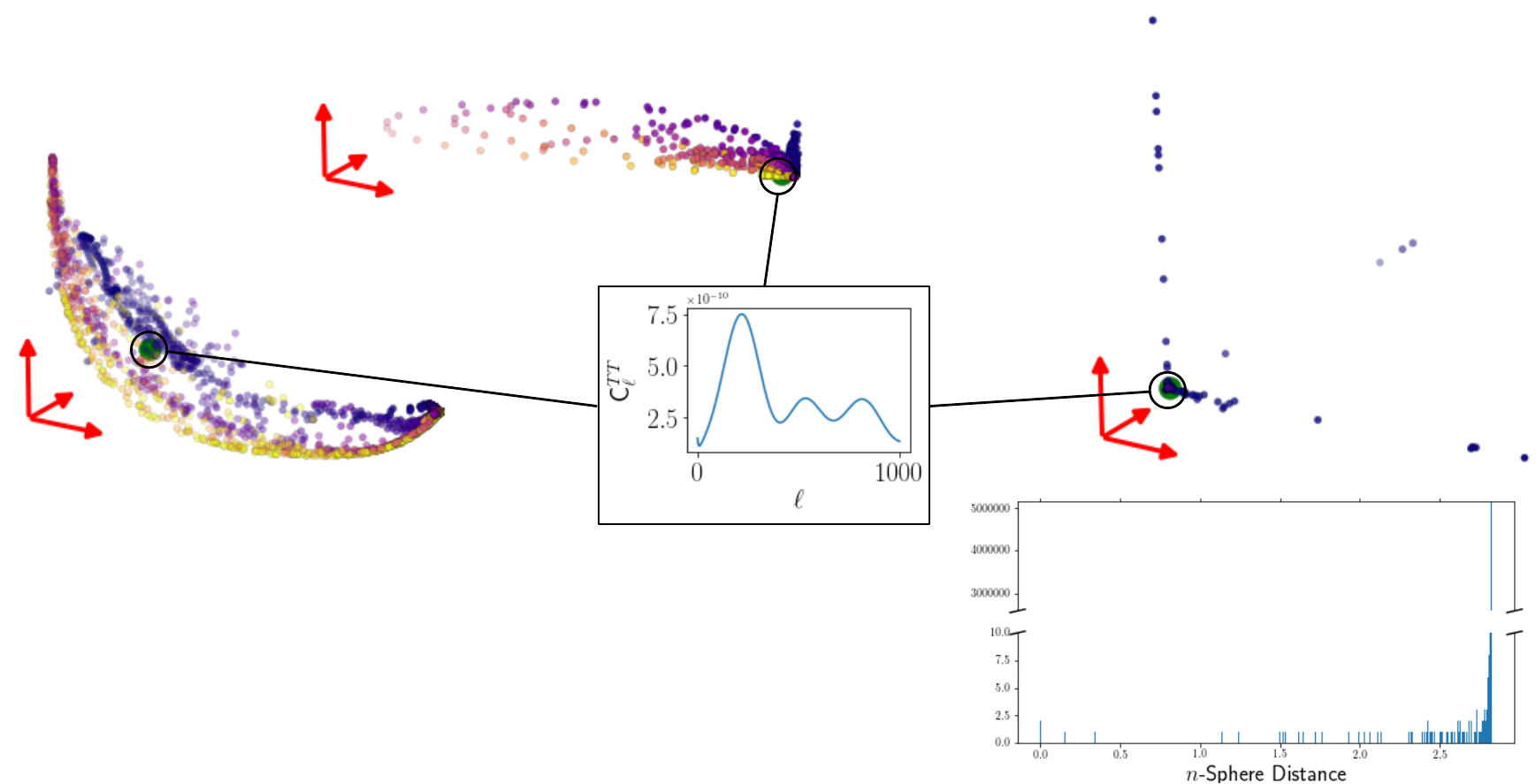
    \caption{{\bf Model manifolds} for the \LCDM\ cosmological model predictions of the cosmic microwave background radiation up to $\ell = 1000$. All manifolds are plotted for the same data, and colored by $A_s$, the primordial fluctuation amplitude. Our universe is indicated in green in all plots. All manifolds are plotted for the first three principal components. (a) The manifold embedded in the $C_\ell$ spectra space, which in non-Euclidean and distorted. (b) The manifold in our intensive embedding, with a histogram of distances between points. The distances are spread out over a wide range, making the visualization useful. (c) The manifold embedded on an $n$-sphere of radius two, where all points become effectively orthogonal. (d) Temperature spectrum for our universe, connected to its location in the three different embedding spaces. (e) Histogram of distances in our intensive embedding space. (f) Histogram of distances in the $n$-sphere embedding space, illustrating that most points are almost a distance $\sqrt{8}$ away, as far apart as possible on the positive orthant on the $n$-sphere of radius two. The distances in the two histograms are related, since the intensive distance $d_I$ can be expressed as a function of the extensive $n$-sphere Helligner distance $d_H$. As a result, minimizing the distance $d_I$ will also minimize $d_H$.}
    \label{fig:CMBManifolds}
\end{figure*}

For regular least-squares fitting, $\Omega$ would simply be a constant matrix representing experimental uncertainty. However, for CMB spectra, it is parameter dependent; it varies with the $C_\ell$. Geometrically, this can be interpreted as the metric in the $C_\ell$ embedding space. Since it varies with the $C_\ell$, it produces a non-Euclidean embedding. Visualizing the model manifold in this space is therefore problematic, since the space is warped and distorted, and distances are not faithfully represented as show in Fig.~\ref{fig:CMBManifolds}~a. This problem can be solved if we instead consider the probability distributions the $C_\ell$ correspond to for different parameters. In Table~\ref{table:CMBparams} we present the range of \LCDM\ parameters explored in our model manifold. In for following section we explore these probability distributions for different $C_\ell$.

\begin{table}[htbp]
\caption{Parameter ranges used to create the model manifolds illustrated in Fig.~\ref{fig:CMBManifolds}. Spectra were generated using CAMB software package.~\cite{Lewis:1999bs} \label{table:CMBparams}}
\begin{ruledtabular}
\begin{tabular}{c l l }
 \textbf{Parameter} & \textbf{Min. Value} & \textbf{Max Value}\\ 
 \hline
        $\tau$ & 0.01 & 0.16\\
        $\eta$ & 0.01 & 0.999 \\
        $A_s$ & $1.0\times 10^{-15}$ & $1.0\times 10^{-7}$ \\
        $h_0$ & 0.091 & 10.0 \\
        $\Omega_bh^2$ & 0.0005 & 99.0 \\
        $\Omega_ch^2$ & 0.0002 & 98.0 \\
\end{tabular}
\end{ruledtabular}
\end{table}

\section{$n$-Sphere isometric embedding and its failings}
\label{sec:HypersphereEmbeddingFailures}
The set of probability distributions from a model generates a `probability simplex', since they are all normalized to one. The Fisher Information Matrix in this space is non-Euclidean; it is diagonal with entries that are parameter dependent. For instance, in this space the Fisher Information for the Ising model is given by
\begin{align}
g_{\mu\nu}(\btheta) &= \sum_{\bS_i} \partial_\mu\Like(\bS_i|\btheta) \partial_\nu\Like(\bS_i|\btheta) \Like(\bS_i|\btheta) \nonumber \\
&= \sum_{\bS_i} \delta_{\bS_i,\bS_\mu}\delta_{\bS_i,\bS_\nu}\Like(\bS_i|\btheta) \nonumber \\
&= \delta_{\bS_\mu,\bS_\nu} \Like(\bS_\mu|\btheta).
\end{align}
This is a non-Euclidean metric, since it is not proportional to $\delta_{\bS_\mu,\bS_\nu}$. It has a parameter dependent component given by the Boltzmann likelihood of being in a given spin state. This is similar to the manifold illustrated in Fig.~\ref{fig:CMBManifolds}~a, which shows the non-Euclidean embedding of CMB spectra in a space whose metric is also point-dependent and detailed in Section~\ref{sec:OmegaMatrix}.

If instead we consider the square root of a normalized probability distribution of a fit to data $\bx$, $2\sqrt{\Like(\bx | \btheta)}$, then we can generate a model manifold embedded on an $n$-sphere of radius two, such that the metric is the Fisher Information Metric. Since $\Like (\bx|\btheta)$ is normalized to one and always positive, the dot product between $\sqrt{\Like(\bx|\btheta_1)}$ and $\sqrt{\Like(\bx|\btheta_2)}$ must be less than one, or equal to one if the likelihood functions are the same. The distance between two points on this $n$-sphere is proportional to the Hellinger distance~\cite{hellinger}, an \textit{f}-divergence similar to the Kullback-Leibler divergence. It is straightforward to show that the metric for this $n$-sphere of radius two embedding is given by the Fisher Information Matrix, by considering the distance for some small perturbation $\delta\btheta$:
\begin{align}
\label{eq:FIM}
&\left<2\sqrt{\Like(\bx | \btheta)}-2\sqrt{\Like(\bx | \btheta+\delta\btheta)},2\sqrt{\Like(\bx | \btheta)}-2\sqrt{\Like(\bx | \btheta+\delta\btheta)}\right> \nonumber \\
 &= 8\left(1-\int\sqrt{\Like(\bx | \btheta)}\sqrt{\Like(\bx | \btheta+\delta\btheta)}\diff\bx\right) \nonumber \\
 &= -4\int\delta\theta^\alpha\partial_\alpha\Like(\bx | \btheta)\diff\bx-\int\frac{\partial_\alpha\partial_\beta\Like(\bx|\btheta)\delta\theta^\alpha\delta\theta^\beta}{\Like(\bx|\btheta)}\diff\bx + \mathcal{O}(\delta\btheta^3)\nonumber \\
 &=\delta\theta^\alpha\delta\theta^\beta\underbrace{\int\partial_\alpha\log\left[\Like(\bx|\btheta)\right]\partial_\beta\log\left[\Like(\bx|\btheta)\right]\Like(\bx|\btheta)\diff\bx}_{\text{Fisher Information Matrix}}
\end{align}

This embedding is therefore isometric, preserving distances as given by the Hellinger divergence~\cite{hellinger}. Unfortunately, there is a maximum distance any two points can be in this embedding, and that is given by the points on the poles attached to the positive orthant of the $n$-sphere. These are a distance $\sqrt{8}$ apart if the radius is two. Therefore, as more and more data are collected, creating increasingly orthogonal points, the manifold `winds around' the $n$-sphere. The image generated by this is not `faithful', in the sense that it does not allow for low-dimensional representations, as shown in Fig.~\ref{fig:CMBManifolds}~c.

\subsection{Cosmic microwave $n$-sphere embedding}
\label{subsec:CMBHypersphere}
There are three important different measures by which the CMB probability distributions can be compared. The first is a scaled Helligner distance~\cite{hellinger} $d_H$ which generates the $n$-sphere embedding (Fig.~\ref{fig:CMBManifolds}~c), the second is our intensive distance $d_I$ shown in (Fig~\ref{fig:CMBManifolds}~b) and the third is the Kullback-Liebler divergence, derived from a normalization of the least squared distance over all possible data that could generate a given set of spectra (which cannot be as easily visualized because it is asymmetric). Note that the manifolds shown in Fig.~\ref{fig:CMBManifolds} are presented with no lensing or $B$ polarization. The full model manifold for all polarization and including lensing, up to $ell=2800$, is insufficiently sampled by the data used in this manuscript and so is not presented.

When embedding CMB predictions on the $n$-sphere of radius two, our squared distance between predictions for parameter sets $\btheta_1$ and $\btheta_2$ is $d_H^2$, the rescaled Hellinger distance, and can be expressed as:
\begin{align}
&d_H^2 = 8\left(1-\int\diff a_{\ell m}\sqrt{p(\{\hat{a}_{\ell m}\}|\btheta_1)}\sqrt{p(\{\hat{a}_{\ell m}\}|\btheta_2)}\right) \nonumber \\
&=8-8\prod_{\ell m}\frac{2^{3/2}}{\left(|C_\ell(\btheta_1)||C_\ell(\btheta_2)||C_\ell(\btheta_1)^{-1} + C_\ell(\btheta_2)^{-1}|^2\right)^{1/4}}
\end{align}
where the last expression is derived by taking a straightforward integral using Eq.~\ref{eq:probCMBDef}. As the expansion is taken to higher order, the product rapidly converges to zero, resulting in a distance of $\sqrt{8}$ for all but very small changes in parameters, as shown in the distance histogram of Fig.~\ref{fig:CMBManifolds}~f which illustrates the huge peak. The model manifold for CMB spectra embedded on the $n$-sphere are represented in Fig.~\ref{fig:CMBManifolds}~c. Our intensive distance, $d_I$, is related to a non-linear function of the Hellinger distance
\begin{align}
d_I^2 &=8\log\left(1-\frac{d_H^2}{8}\right) \nonumber \\
&= -8 \sum_{\ell} \frac{2\ell +1}{4}\log\left(\frac{|C_\ell(\btheta_1) + C_\ell(\btheta_2)|^2}{64|C_\ell(\btheta_1)||C_\ell(\btheta_2)|}\right).
\end{align}

It would be natural from an information geometry point of view to minimize $d_I^2$ in finding best fits of the model to data. In practice, the astrophysics community minimizes using a least squares method normalized over all possible data that could yield the same results~\cite{PlanckLikelihood,CMBHamimeche,AmariInfo}. One can easily show that this is in fact the Kullback-Liebler divergence and is expressed as
\begin{align}
-\frac{1}{2}\sum_\ell(2\ell+1)\left(\text{Tr}\left(\hat{C}_\ell C_\ell(\btheta)^{-1}\right)+\log\frac{|C_\ell(\btheta)|}{|\hat{C}_\ell|} - 3\right)
\end{align}
where $\hat{C}_\ell$ represents the measured CMB spectra from experimental data, and $C_\ell(\btheta)$ are the spectra predicted for parameters $\btheta$. In both cases, the Kullback-Leibler and the Hellinger divergences measure distance between probability distributions. These two divergences belong to a broader class of $f$-divergences, where $f$ is a convex function. In all these cases, the distance between two probability distributions is characterized by the choice of $f$ function, and the metric (the divergence between nearby parameter sets) is proportional to the Fisher Information Matrix~\cite{Amari}.

\section{Intensive manifold as an isometric embedding}
\label{sec:IntensiveIsometricity}
Distances between predictions for two parameter combinations $\btheta_1$ and $\btheta_2$ in our intensive embedding are given by:
\begin{equation}
d_I^2(\btheta_1,\btheta_2) = -8\log\left<\sqrt{\Like(\bx|\btheta_1)},\sqrt{\Like(\bx|\btheta_2)}\right>
\end{equation}
To determine the metric for this embedding, we consider a small parameter perturbation $\delta\btheta$ around some parameter combination $\btheta$:
\begin{align}
d_I^2(\btheta,\btheta+\delta\btheta) &= -8\log\int\diff\bx\sqrt{\Like(\bx|\btheta)}\sqrt{\Like(\bx|\btheta+\delta\btheta)} \nonumber \\
&= \int\diff\bx\frac{\partial_\alpha\Like(\bx|\btheta)\partial_\beta\Like(\bx|\btheta)}{\Like(\bx|\btheta)}\delta\theta^\alpha\delta\theta^\beta + \mathcal{O}(\delta\btheta^3)
\end{align}
producing the same fisher information metric shown in Eq.~\ref{eq:FIM}. For simplicity, we have dropped all terms from the expansion equal to zero. By preserving the local metric, our intensive embedding is isometric.

\section{Principal component analysis for the intensive manifold embedding}
\label{sec:IntensivePCA}

In order to visualize the various model manifolds in Fig.~\ref{fig:CMBManifolds}, we performed a principal component analysis on the data. This process rotates the data into an orthogonal basis such that the first component is along the direction of greatest variation, the second direction is along the direction of second greatest variation, and so on. In order to accomplish this, a data set $\{\bd^{(J)}\}$ is produced, indexed by the superscript $(J)$. A data matrix can be produced, $D_{iJ} = d^{(J)}_i$, and the index $i$ indicates the vector component of the data. This can be discrete, such as the probability state vector for the Ising model, or continuous such as the likelihood of observing a certain fluctuation $a_{lm}$ in a CMB map.

The columns of $D$ are centered, producing a matrix $M_{iJ} = d^{(J)}_i - \tilde{d}_i$ where $\tilde{\bd} = \frac{1}{n}\sum_J \bd^{(J)}$ and $n$ is the number of data points. A singular value decomposition is normally performed on $M = U\Sigma V^T$, and the $i$th principal component for data point $J$ is now given by $\Sigma_{ii}U_{Ji}$. This works for discrete data, but in the case of continuous data (such as likelihood functions) we must take a slightly different approach.

We construct a matrix of dot products,
\begin{align}
(MM^T)_{JK} &= \sum_i (d^{(J)}_i - \tilde{d}_i)(d^{(K)}_i - \tilde{d}_i) \nonumber \\
&= \bd^{(J)}\cdot \bd^{(K)} - \left(\bd^{(J)}\cdot\tilde{\bd} + \bd^{(K)}\tilde{\bd}\right) + \tilde{\bd}\cdot\tilde{\bd} \nonumber \\
&= \bd^{(J)}\cdot \bd^{(K)} -\frac{1}{n}\sum_L\left(\bd^{(J)}\cdot \bd^{(L)} + \bd^{(K)}\cdot \bd^{(L)}\right)  \nonumber \\
&+ \frac{1}{n^2}\sum_{L,L'} \bd^{(L)}\cdot \bd^{(L')}.
\end{align}
Since this matrix can also be expressed as $MM^T = U\Sigma V^T V \Sigma U^T = U\Sigma^2 U^T$, we can find the principal components of our data by finding the eigenvalues and eigenvectors of the matrix of dot products. If we consider the case where data points $\bd^{(J)}$ are the square roots of the probability distribution predicted from parameter combination $\btheta_J$ then, using the dot product for our $N$ replicated system, we can write out the components of $(MM^T)$ as:
\begin{align}
(MM^T)_{JK} =& \frac{\left<\btheta_J;\btheta_K\right>^N}{N} - \sum_L\left(\frac{\left<\btheta_J;\btheta_L\right>^N}{nN} + \frac{\left<\btheta_K;\btheta_L\right>^N}{nN}\right) \nonumber \\
&+ \sum_{L,L'}\frac{\left<\btheta_L;\btheta_{L'}\right>^N}{n^2N}
\end{align}
In the limit where the number of replicas go to zero, the matrix of dot products becomes:
\begin{align}
(MM^T)_{JK} =& \log\left<\btheta_J;\btheta_K\right> \nonumber \\
&- \frac{1}{n}\sum_L\left(\log\left<\btheta_J;\btheta_L\right> + \log\left<\btheta_K;\btheta_L\right>\right) \nonumber \\
&+ \frac{1}{n^2}\sum_{L,L'}\log\left<\btheta_L;\btheta_{L'}\right>
\label{eq:MatrixDots}
\end{align}

In the case where the number of replicas is not a whole number, and in the limit where it tends to zero, the matrix of dot products is no longer positive definite. As a result, we can obtain positive and negative eigenvalues from its decomposition, leading to real and imaginary principal components.

In order to include experimental data into this plot, a probability distribution from the data must be generated. The dot product between these measurements and predicted distributions can then be calculated, and the results used in the matrix of dot products in Eq.~\ref{eq:MatrixDots} to find it's projection along the principal components. The distance from the point to the manifold is given by our intensive distance from the point to the best fit.

\bibliography{IntensiveManifold}